\documentclass{elsart}
\usepackage[]{amsmath,amssymb}
\usepackage[]{epsfig}
\begin{document}
\begin{frontmatter}
\begin{flushright}
\texttt{IZTECH-P-10/06\\ICCUB-10-200}
\end{flushright}%
\title{\bf Higgs-Palatini Inflation and Unitarity}
\author[Barcelona]{Florian Bauer}
\ead{fbauerphysik@eml.cc}
\and
\author[Izmir]{Durmu{\c s} A. Demir}
\ead{demir@physics.iztech.edu.tr}
\address[Barcelona]{High Energy Physics Group, Dept.\ ECM,
Institut de Ci\`encies del Cosmos~(ICC)\\
Universitat de Barcelona, Mart\'i i Franqu\`es 1,
08028 Barcelona, Catalonia, Spain}
\address[Izmir]{Department of Physics, {\.I}zmir Institute of Technology, TR35430, {\.I}zmir, Turkey}
\begin{abstract}
In the Higgs inflation scenario the Higgs field is strongly coupled
to the Ricci scalar in order to drive primordial inflation. However,
in its original form in pure metric formulation of gravity, the ultraviolet~(UV) cutoff of the Higgs interactions and the Hubble rate are of the same magnitude, and
this makes the whole inflationary evolution dependent of the unknown UV
completion of the Higgs sector. This problem, the unitarity violation,
plagues the Higgs inflation scenario. In this letter we show that, in the Palatini formulation of gravitation, Higgs inflation does not suffer from unitarity
violation since the UV cutoff lies parametrically much higher than
the Hubble rate so that unknown UV physics does not disrupt the inflationary
dynamics. Higgs-Palatini inflation, as we call it,  is, therefore,
UV-safe, minimal and endowed with predictive power.
\end{abstract}
\end{frontmatter}
\section{Introduction}
\label{sec:Intro}
In general, the negative pressure required by successful inflation is readily provided by a slowly varying
scalar field. As a conservative scenario, this scalar field -- the inflaton -- may be identified with
the Higgs boson of the standard electroweak theory (SM)~\cite{HiggsInf-001-0710}. This approach thus
exalts the Higgs boson to the level of `god particle' with which particle physics and cosmology become
intimately related.

Higgs inflation rests on a non-minimal ($\zeta \neq 0$), non-conformal ($\zeta \neq 1/6$) interaction~$\zeta\phi^{2}R$
between the Higgs field $\phi$ and  space-time's scalar curvature $R$. Typically, $\zeta \gg 1$~\cite{HiggsInf-001-0710}. Recently,
this approach has been under dense discussion in regard to unitarity violation ~\cite{UV-Theory-001-0902,Lerner:2009na,Unitarity-001-1008,AsympSafety-1011}. Indeed, it has been argued that the Higgs inflation setup necessitates an UV completion at the scale $M_{\text{Pl}}/\zeta$, where $M_{\text{Pl}}$ is the Planck mass. This scale is close to the energy scale of inflation.  The essence of the problem is that new degrees of freedom  with masses around $M_{\text{Pl}}/\zeta$ can exist, and they prohibit making reliable predictions on the inflationary dynamics. Indeed, any prediction will exhibit a strong dependence on unknown physics above $M_{\text{Pl}}/\zeta$~\cite{UV-Theory-001-0902,UV-Theory-002-0903}, and in the absence of a detailed UV-completion, the inflationary evolution cannot be taken under control. Various aspects of inflation with non-minimally coupled scalars have been analyzed in~\cite{NonMinInfl-006-9505,NonMinInfl-003-0911,NonMinInfl-004-1002,NonMinInfl-001-1002,NonMinInfl-005-1003,NonMinInfl-002-1006}.

All the properties of Higgs inflation, studied in~\cite{HiggsInf-001-0710,UV-Theory-001-0902,Unitarity-001-1008, AsympSafety-1011,UV-Theory-002-0903,NonMinInfl-003-0911,NonMinInfl-004-1002,NonMinInfl-001-1002,NonMinInfl-005-1003,NonMinInfl-002-1006}, are akin to purely metrical gravity, that is, General Relativity. In this formulation, the affine connection {\it i.\ e.{}} the Levi-Civita connection is fully expressed in terms of the metric tensor from the scratch, and hence, the entire geometry is governed by the metric tensor.  This formulation, the so-called {\it second order formalism}, suffers from the need to an extrinsic curvature contribution for the Einstein-Hilbert action to reproduce the Einstein field equations of gravitation.

In this work, we shall analyze Higgs inflation and the associated unitarity violation in the Palatini formulation~\cite{palatini} of gravitation. This alternative formulation stems from the fact that the affine connection and the metric tensor are {\it a priori} independent geometrical variables, and if they are to exhibit any relationship it must arise from dynamical equations {\it a posteriori}. This very setup, the so-called
{\it Palatini formulation} or first-order formalism, does not necessarily admit the Levi-Civita connection if the matter sector depends explicitly on the affine connection~\cite{differ,differ2}. In this formalism, connection and metric are independent geometrodynamical variables~\cite{Palatini-HI-0803}, and this very fact has physically interesting consequences for inflation~\cite{Palatini-HI-0803,Palatini-002-0801,Palatini-003-0804,Palatini-007-0805,Palatini-001-1010,Palatini-004-1010} as well as other phenomena like the cosmological constant problem~\cite{Palatini-005-0910,Palatini-006-1007}.

In the body of the work,  we shall show that, {\it Higgs inflation in the Palatini formulation is not only natural as was explicitly shown in~\cite{Palatini-HI-0803} but also unitary since the UV  cutoff scale~$M_{\text{\upshape Pl}}/\sqrt{\zeta}$ lies much higher than the inflationary scale~$H\sim M_{\text{\upshape Pl}}/\zeta$}. This level splitting is wide enough to allow for making reliable predictions about the inflationary dynamics since unknown UV physics lies too high to leave a discernible impact. This result constitutes another striking difference between the metric and Palatini formulations in regard to the ones given in~\cite{Palatini-HI-0803}.

In Sec.~\ref{sec:HiggsInfl} below we discuss Higgs inflation in the Palatini formalism. This section mainly parallels~\cite{Palatini-HI-0803} except that the scalar field~$\phi$ is now gauged under the electroweak gauge group. In Sec.~\ref{sec:UVCutoff} we compute the UV cutoff scale. Therein, we comparatively analyze it  in the metric and the Palatini formulation. In Sec.~\ref{sec:Conclusions} we conclude.

\section{Higgs Inflation}
\label{sec:HiggsInfl}
The Higgs inflation scenario is based on the Jordan frame action
\begin{eqnarray}
\mathcal{S}=\int d^{4}x\sqrt{|g|}\left[- \frac{1}{2} {\mathcal{M}}^{2}({\mathcal{H}}) g^{\mu\nu}\mathbb{R}_{\mu\nu}\left(\Gamma\right)+\frac{1}{2}g^{\mu\nu}(D_{\mu} {\mathcal{H}})^{\dagger}(D_{\nu} {\mathcal{H}})-V({\mathcal{H}})\right]
\label{eq:Action-Jordan}
\end{eqnarray}
where ${\mathcal{H}}$ is the SM Higgs field (an $SU(2)_L$ doublet with non-vanishing hypercharge). The non-minimal Higgs-curvature coupling is contained in
\begin{eqnarray}
{\mathcal{M}}^{2}({\mathcal{H}}) = {M^{2}+\zeta {\mathcal{H}}^{\dagger} {\mathcal{H}} }
\end{eqnarray}
where $M$ is a mass scale to be eventually related to the gravitational coupling $M_{\text{Pl}}$.  The Higgs potential energy density
is the usual quartic one
\begin{eqnarray}
V({\mathcal{H}}) = \frac{1}{4} \lambda \left( {\mathcal{H}}^{\dagger} {\mathcal{H}}  - v^2  \right)^2
\end{eqnarray}
where $v$ -- the Higgs vacuum expectation value (VEV) -- approximately equals the top quark mass $m_t$. The quartic coupling $\lambda$ is a small fraction of unity.

In pure metric theory, the connection $\Gamma^{\lambda}_{\alpha\beta}$ in (\ref{eq:Action-Jordan}) is fixed to the Levi-Civita connection
\begin{eqnarray}
\widehat{\Gamma}^{\lambda}_{\alpha\beta} = \frac{1}{2} g^{\lambda \rho}\left( \partial_{\alpha} g_{\beta \rho} + \partial_{\beta} g_{\rho \alpha} - \partial_{\rho} g_{\alpha\beta} \right)
\end{eqnarray}
so that $\mathbb{R}_{\mu\nu}\left(\widehat{\Gamma}\right)$ is directly given in terms of the metric tensor. Therefore, in pure metrical gravity the metric tensor determines the Riemann and Ricci tensors themselves via the Levi-Civita connection. Furthermore, it determines the Ricci scalar by contracting the Ricci tensor. Consequently, a transformation of the metric tensor such as
\begin{eqnarray}
g_{\mu\nu}\rightarrow e^{-2\omega} g_{\mu\nu} \equiv    \frac{M_{\text{Pl}}^{2}}{{\mathcal{M}}^{2}({\mathcal{H}})} g_{\mu\nu}\label{eq:ConfTrafo}
\end{eqnarray}
modifies both the Ricci tensor and Ricci scalar.

In the Palatini formalism, where connection and metric are fundamentally independent geometrical variables, a transformation of the metric does induce no transformation on the connection and vice versa. In fact, $\Gamma^{\lambda}_{\alpha\beta}$, bearing no relation to the Levi-Civita connection  $\widehat{\Gamma}^{\lambda}_{\alpha\beta}$, does exhibit no change at all while the metric transforms as in~(\ref{eq:ConfTrafo}). Therefore, there arise striking differences between the two formalisms in regard to the
transformation of the action~(\ref{eq:Action-Jordan}) under~(\ref{eq:ConfTrafo}):
\begin{eqnarray}
\mathcal{S} & = & \int d^{4}x\sqrt{|g|}\left[-\frac{M_{\text{Pl}}^{2}}{2}g^{\mu\nu}\mathbb{R}_{\mu\nu}\left(\widehat{\Gamma}\right) - e^{-4\omega}\,V({\mathcal{H}}) \right.\nonumber \\
 & + & \left.\frac{1}{2}e^{-2\omega}(D_{\mu}{\mathcal{H}})^{\dagger}(D^{\mu}{\mathcal{H}})+f\cdot\frac{1}{2}e^{-4\omega}\frac{3\zeta^{2}}{2 M_{\text{Pl}}^{2}} \partial_{\mu}\left( {\mathcal{H}}^{\dagger} {\mathcal{H}}\right)   \partial^{\mu}\left( {\mathcal{H}}^{\dagger} {\mathcal{H}}\right) \right]\label{eq:Action-Einstein}
\end{eqnarray}
where
\begin{eqnarray}
f = \left\{\begin{array}{l} 1\;\;\;\;\; \mbox{in\ metric\ formalism},\\  0\;\;\;\;\; \mbox{in\ Palatini\ formalism}.\end{array}\right.
\end{eqnarray}
It is clear that, in the Einstein frame action (\ref{eq:Action-Einstein}),  the connection that generates the Ricci tensor $\mathbb{R}_{\mu \nu}$ is always the Levi-Civita connection $\widehat{\Gamma}^{\lambda}_{\alpha\beta}$ since metric and Palatini formalisms turn out to be dynamically equivalent when the Einstein-Hilbert term is not multiplied by a function of $\mathcal{H}$ as in the Jordan frame action (\ref{eq:Action-Jordan}), and when the other sectors especially the matter sector do not involve the connection $\Gamma^{\lambda}_{\alpha\beta}$, explicitly~\cite{Palatini-HI-0803}.

The Einstein frame action (\ref{eq:Action-Einstein}) manifestly reveals how unitarity problems occur in Higgs inflation. It is convenient to discuss first the~$f=1$ case corresponding to metrical gravity. The term proportional to~$f$ is a dimension-6 operator, and for small Higgs field values
\begin{eqnarray}
\label{UVmetric}
\Lambda_{\text{(metric)}} \equiv \frac{M_{\text{Pl}}}{\zeta\sqrt{6}}
\end{eqnarray}
behaves as the UV cutoff. This scale lies significantly below $ {M_{\text{Pl}}}$ for large $\zeta$. To avoid problems with unitarity, an UV completion must be introduced at energies around~$\Lambda_{\text{(metric)}}$. Besides, according to \cite{Palatini-HI-0803}, the Friedmann equation in the Einstein frame, during slow-roll inflation, reads to be
\begin{eqnarray}
\frac{3 M_{\text{Pl}}^{2}}{8\pi} H^{2}=V({\mathcal{H}}) \simeq \frac{M_{\text{Pl}}^{4}\lambda}{4\zeta^{2}}
\end{eqnarray}
in both metric and Palatini formalisms. This equality thus fixes the characteristic energy
scale of the inflationary dynamics to be the Hubble rate
\begin{eqnarray}
\label{hubble}
H=\frac{M_{\text{Pl}}}{\zeta}\sqrt{\frac{4\pi\lambda}{6}}\simeq \Lambda_{\text{(metric)}} \label{eq:HubbleEinstein}
\end{eqnarray}
when $\lambda$ is ${\cal{O}}(0.1)$. What this equality is showing is that the inflationary dynamics occurs energetically close to the energy scale~$\Lambda_{\text{(metric)}}$, where the UV-safe theory has been implemented. Hence, the whole inflationary epoch becomes highly sensitive to `unknown physics' at~$\Lambda_{\text{(metric)}}$. In fact, this `unknown physics' should be capable of rehabilitating the Higgs sector all the way up to ${M_{\text{Pl}}}$ by embedding it into a more fundamental theory. It is clear that for it to accomplish this, most probably, new particles will be needed to populate  the energy range from  $\Lambda_{\text{(metric)}}$ up to  ${M_{\text{Pl}}}$  with appropriate dynamics. These new ingredients will surely affect the Higgs potential and the inflationary evolution it drives~\cite{unitarity-burgess,UnitaryByHand-1005,Giudice2010-1010}.

One notes that, for large Higgs values during inflation, the action (\ref{eq:Action-Einstein}) implies an UV cutoff around the Planck mass. Nevertheless, inflation is influenced by the (unknown) UV-complete theory to be implemented at the scale~$\Lambda_{\text{(metric)}}\ll M_\text{Pl}$ in order to unitarize the model for small Higgs values. Therefore, the entire problem of unitarity violation, as it arises from the dimension-6 operator in  (\ref{eq:Action-Einstein}), is akin to the metric formalism. Indeed, the operator in question is completely  absent in the Palatini formalism for which $f=0$.  As will be detailed in the next section, this is one of the most advantageous aspects of the Palatini formalism in regard to unitarity violation.

\section{The UV Cutoff}
\label{sec:UVCutoff}
In this section we shall explicitly compute the UV cutoffs $\Lambda_{\text{(metric)}}$ and $\Lambda_{\text{(Palatini)}}$ in the two formalisms.  For energies close to this cutoff $\Lambda_\text{UV}$ the perturbative analysis breaks down due to strong coupling effects or unitarity violation. Therefore, one has to have a clear understanding
of the UV completion in order to make reliable predictions about different stages of the inflationary epoch. The brief discussion in
the last section should have made it clear that,  the decisive quantity is the relative size of the Hubble rate and $\Lambda_{\text{UV}}$: If they are of similar size the inflaton energetically sits at $\Lambda_{\text{UV}}$ and unitarity violation occurs.  If the two scales are sufficiently split {\it i.\ e.{}} $\Lambda_{\text{UV}}\gg H$ then  unitarity violation is avoided so that the properties of the inflationary phase are determined solely by the seed model in (\ref{eq:Action-Jordan}), and no knowledge of the UV theory is needed. In such a case, Higgs inflation would be minimal and predictive.

In general, the SM Higgs field
\begin{eqnarray}
{\mathcal{H}}\equiv \left(\begin{array}{c} \varphi^{+}\\ \phi + i \varphi^0\end{array}\right)
\end{eqnarray}
is composed of the Higgs boson $\phi$, charged Goldstone boson $\varphi^{+}$,
and neutral Goldstone boson $\varphi^0$. In unitary gauge, wherein $\varphi^{+}$ and
$\varphi^0$ are respectively swallowed by the $W^{+}$ and $Z$ bosons, the Higgs field
reduces to $\left(0, \phi\right)^{T}$. Then, for the sake of simplicity, specializing to an
Abelian gauge group (like the hypercharge gauge group in the SM) with gauge boson $A_{\mu}$
and gauge coupling $g$, the Higgs kinetic term takes the form
\begin{eqnarray}
g^{\mu\nu}(D_{\mu}{\mathcal{H}})^{\dagger}(D_{\nu}{\mathcal{H}})=g^{\mu\nu}\left[(\partial_{\mu}\phi)(\partial_{\nu}\phi)+g^{2}\phi^{2} A_{\mu}A_{\nu}\right] \label{eq:HiggsKinTerm}
\end{eqnarray}
where the second term at the right-hand side generates the gauge boson mass-squared $2 g^2 v^2$ upon spontaneous breakdown of the gauge symmetry by the nonvanishing VEV $\langle \phi \rangle = v$.

The goal of the analysis here is to determine the scale of the unitarity violation, if any, in the Einstein frame. For this purpose,
it is necessary to identify the potential sources, and analyze them in regard to their UV boundary.  In the Einstein frame action~(\ref{eq:Action-Einstein}), the reduction of the Higgs kinetic term via~(\ref{eq:HiggsKinTerm})  gives rise to the Higgs-gauge interaction
\begin{eqnarray}
\mathcal{U}_{g} \equiv  \frac{M_{\text{Pl}}^{2}}{{\mathcal{M}}^{2}(\phi)} \left(g^{2} \phi^2 A_{\mu}A^{\mu}\right) \label{eq:a1}
\end{eqnarray}
from which the cutoff $\Lambda_{\text{UV}}$ can be read off in both formalisms.

Apart from the gauge-Higgs interaction in (\ref{eq:a1}), there are also Higgs-fermion interactions coming from the Yukawa interactions
$\overline{\Psi}\Psi\phi$ in the Jordan frame. %
This interaction term becomes
$e^{-4\omega}\overline{\Psi}\Psi\phi$ in the Einstein frame. Moreover,  fermions attain their minimal
kinetic terms after the rescaling  $\Psi\rightarrow e^{(3/2)\omega}\Psi$. Hence, one arrives at
\begin{eqnarray}
{\mathcal{U}}_{f}\equiv  \frac{M_{\text{Pl}}}{\sqrt{{\mathcal{M}}^{2}(\phi)}}
\overline{\Psi}\Psi\phi \label{eq:a2}
\end{eqnarray}
as another higher-order interaction to be examined for determining the relevant cutoff $\Lambda_{\text{UV}}$.

Finally, Higgs-Higgs interactions can also bring unitarity violation through
\begin{eqnarray}
\mathcal{U}_{h}\equiv  \frac{M_{\text{Pl}}^{4}}{\left({\mathcal{M}}^{2}(\phi)\right)^2} V(\phi)\label{eq:a3}
\end{eqnarray}
which is nothing but the Higgs potential in the Einstein frame. The quantities $\mathcal{U}_g$, $\mathcal{U}_f$ and $\mathcal{U}_h$ represent three distinct classes of higher-dimensional operators which can exhibit unitarity violation in the inflationary epoch. They each will be examined below separately and comparatively in the two formalisms.

\subsection{Metric formalism}
\label{sub:Metric}
Unitarity violation in metrical Higgs inflation has been extensively analyzed in literature~\cite{unitarity-burgess,UnitaryByHand-1005,Giudice2010-1010}.
Here we shall content with the determination of $\Lambda_{\text{(metric)}}$ by considering only the gauge-Higgs contribution $\mathcal{U}_{g}$.
The other channels, $\mathcal{U}_f$ and $\mathcal{U}_h$, can be analyzed similarly.

If $\psi$ is to be the Higgs field with a canonical kinetic term, it has to be related to $\phi$ by~\cite{Palatini-HI-0803}
\begin{eqnarray}
\frac{d\psi}{d\phi}=\frac{M_{\text{Pl}}}{M}\frac{\sqrt{1+\frac{\phi^{2}}{M^{2}}(6\zeta^{2}+\zeta)}}{1+\zeta\frac{\phi^{2}}{M^{2}}},\label{eq:dTrafo-Metric}
\end{eqnarray}
integration of which yields the requisite functional relation between $\psi$ and $\phi$.

For low values of the Higgs field $\phi$ {\it i.\ e.{}} $\sqrt{\zeta}\phi \ll M$,  equation (\ref{eq:dTrafo-Metric}) integrates to give the power series
\begin{eqnarray}
\psi=\frac{M_{\text{Pl}}}{\zeta} \left[ \frac{\zeta \phi}{M} + \left(\frac{\zeta\phi}{M}\right)^{3}+\dots\right]
\end{eqnarray}
which can be inverted to obtain
\begin{eqnarray}
\phi=\frac{M \psi}{M_{\text{Pl}}} - \frac{\zeta^2}{M^2} \left(\frac{M \psi}{M_{\text{Pl}}}\right)^3 + \dots
\end{eqnarray}
where $\dots$ stand for progressively higher order terms in the expansion. Using this expression for $\phi$, the
Higgs-gauge interaction contribution $\mathcal{U}_g$ in (\ref{eq:a1}) becomes
\begin{eqnarray}
\mathcal{U}_g &=& \frac{M_{\text{Pl}}^{2}}{M^2} g^{2}   A_{\mu}A^{\mu} {\phi^{2}} \left[1 - \frac{\zeta\phi^{2}}{M^{2}}+\dots\right]\nonumber\\
&=& g^{2} A_{\mu} A^{\mu} \psi^{2} \left[1 - \frac{1}{3 \Lambda^2_{\text{(metric)}}} \left(1 + \frac{1}{2 \zeta}\right)
\psi^{2}+\dots\right]
\end{eqnarray}
from which it is clear that $\Lambda_{\text{(metric)}}$ defined in (\ref{UVmetric}) is indeed the UV cutoff
for higher-dimensional interactions between the Higgs and gauge fields. Therefore, the Higgs inflation model
can work safely only at energies sufficiently below $\Lambda_{\text{(metric)}}$ since the effective field
theory description breaks down as the UV cutoff is approached.

For large values of the Higgs field {\it i.\ e.{}} $\zeta \phi \gg M$, the relation~(\ref{eq:dTrafo-Metric})
integrates to give the exact relation
\begin{eqnarray}
\phi=\frac{M}{\sqrt{\zeta}}\exp\left(\frac{\psi}{\sqrt{6}M_{\text{Pl}}}\right)
\end{eqnarray}
thanks to which the Higgs-gauge coupling term~$\mathcal{U}_g$ can be written as
\begin{eqnarray}
\mathcal{U}_g = 6 \zeta \Lambda^2_{\text{(metric)}} g^{2} A_{\mu} A^{\mu} \left( 1 + \exp{\left(-2 \frac{\psi}{\sqrt{6} M_{\text{Pl}}}\right)} \right)^{-1} \end{eqnarray}
whose dependence on the canonical Higgs field $\psi$ goes through
the ratio $\psi/(\sqrt{6} M_{\text{Pl}})$ with no involvement of $\zeta$. This
clearly shows that the UV cutoff is equal to $\sqrt{6} M_{\text{Pl}}$, which lies
slightly above $M_{\text{Pl}}$.

In summary, in purely metric gravity:
\begin{itemize}
\item For $\zeta \phi \ll M$, the UV cutoff equals $\Lambda_{\text{(metric)}} = M_{\text{Pl}}/(\zeta\sqrt{6})$.
This is the energy scale where the Higgs sector is to be embedded into an UV-safe theory whose degrees of freedom should have masses around $\Lambda_{\text{(metric)}}$. Since the Hubble rate~(\ref{eq:HubbleEinstein}) during inflation is of similar magnitude, these extra fields interfere with the inflaton dynamics. Even if inflation still holds, its initial conditions can be overly sensitive to the UV-completion, as exemplified by \cite{Giudice2010-1010}.

\item For $\zeta \phi \gg M$, the UV cutoff equals $\Lambda_{\text{(metric)}} = \sqrt{6} M_{\text{Pl}}$ itself, which lies safely above the inflationary Hubble rate. However, for the Higgs model to be predictive one has to know the UV theory mentioned above. Apart from that, for $\psi$ values near or above $\Lambda_{\text{(metric)}}$ there arises the well-known naturalness problem of inflation in that the inflaton takes values inside the Planckian territory~\cite{natural,Palatini-HI-0803}.

\item The values of the UV cutoff for small and large field regimes suggest the interesting parametric relation
\begin{eqnarray}
\label{relation}
\Lambda_{\text{(metric)}}(\text{small field},\,\zeta = 1/6) = \Lambda_{\text{(metric)}}(\text{large field})
\end{eqnarray}
so that $\zeta=1/6$ turns out to be a special point. Notably, for this particular value of $\zeta$ the geometrical sector of
the Jordan frame action (\ref{eq:Action-Jordan}) exhibits exact conformal invariance \cite{conformal}. Nonetheless, the
relation (\ref{relation}) between the two UV cutoffs neither holds nor is relevant for a successful Higgs inflation. Indeed, Higgs inflation in 
metrical gravity typically requires $\zeta\simeq 10^{4}$.

\end{itemize}

\subsection{Palatini formalism}
\label{sub:Palatini}
In the Palatini formalism,  the dimension-6 operator in the
Einstein frame action~(\ref{eq:Action-Einstein}) drops out
since the Ricci tensor does not change under the conformal
transformation~(\ref{eq:ConfTrafo}). This is expressed
simply by taking $f=0$ in~(\ref{eq:Action-Einstein}). In this
formalism, the canonically-normalized Higgs field $\psi$ and the original
Higgs field $\phi$ are related via
\begin{eqnarray}
\frac{d\psi}{d\phi}=\frac{M_{\text{Pl}}}{M}\sqrt{\frac{1}{1+\zeta\frac{\phi^{2}}{M^{2}}}},
\end{eqnarray}
integration of which yields
\begin{eqnarray}
\label{rel-palatini}
\phi=\frac{M}{\sqrt{\zeta}}\sinh\left(\frac{\psi\sqrt{\zeta}}{M_{\text{Pl}}}\right)\,.
\end{eqnarray}
In the light of this relation we shall now investigate the problem of unitarity violation
in the Palatini formalism. For this purpose, below we shall compute the UV cutoffs revealed by
the higher-dimensional operators in $\mathcal{U}_g$, $\mathcal{U}_f$ and
$\mathcal{U}_h$.

It is convenient to start the analysis with the Higgs-Higgs interactions encoded in $\mathcal{U}_h$.
After using (\ref{rel-palatini}) in (\ref{eq:a3}), $\mathcal{U}_h$ takes the form
\begin{eqnarray}
\mathcal{U}_h= 9 \lambda \zeta^2 \Lambda_{\text{(metric)}}^4
\frac{\left(\sinh^{2}\left(\frac{\sqrt{\zeta}\psi}{M_{\text{Pl}}}\right) - \frac{\zeta v^{2}}{M^{2}}\right)^{2}}{\left(1+\sinh^{2}\left(\frac{\sqrt{\zeta}\psi}{M_{\text{Pl}}}\right)
\right)^{2}}\label{eq:a3-Pala-exact}
\end{eqnarray}
from which it follows that the UV cutoff is
\begin{eqnarray}
\label{UV-scales}
\Lambda_{\text{(Palatini)}} \equiv \frac{M_{\text{Pl}}}{\sqrt{\zeta}} = \sqrt{6\zeta} \Lambda_{\text{(metric)}}
\end{eqnarray}
One readily observes that for the conformal value of the Higgs-curvature coupling the UV scales of the
two formalisms  become identical {\it i.\ e.{}}  $\Lambda_{\text{(Palatini)}} = \Lambda_{\text{(metric)}} = \sqrt{6}M_{\text{Pl}}$
for $\zeta=1/6$. Of course, this particular relation is irrelevant for Higgs inflation wherein $\zeta$ takes large
values in both formalisms.

The very result that the UV cutoff is ${M_{\text{Pl}}}/{\sqrt{\zeta}}$ holds for both small and large
values of $\psi$. Indeed, in both limits the argument of $\sinh$ function in (\ref{eq:a3-Pala-exact}) remains
unchanged. For a more detailed analysis, however, one can perform the expansion of $\psi$
\begin{eqnarray}
\psi = \psi_0 + h
\end{eqnarray}
about its vacuum expectation value
\begin{eqnarray}
\psi_0 = \frac{M_{\text{Pl}}}{\sqrt{\zeta}}\,\text{arsinh} \left(\frac{\sqrt{\zeta} v}{M} \right)
\end{eqnarray}
which approximately equals ${M_{\text{Pl}} v}/{M}$. Then, expanding (\ref{eq:a3-Pala-exact}) in powers of $h$ one finds
\begin{eqnarray}
\mathcal{U}_h = \lambda \left( v^2 h^2 + v h^3 + \frac{1}{4} h^4 \right) + \dots
\end{eqnarray}
where $\dots$ denote higher order terms suppressed by inverse powers of $\Lambda_{\text{(Palatini)}}$. This
result is nothing but the SM Higgs potential after electroweak breaking.

Having investigated the Higgs potential, we now turn to an analysis of the Higgs-gauge contribution
$\mathcal{U}_g$. Using (\ref{rel-palatini}) in (\ref{eq:a1}) one finds
\begin{eqnarray}
\mathcal{U}_g = g^{2}A_{\mu} A^{\mu} \Lambda_{\text{(Palatini)}}^2 \tanh^{2} \left(\frac{\sqrt{\zeta}\psi}{M_{\text{Pl}}}\right) \label{eq:a1-Pala-exact}
\end{eqnarray}
which is manifestly seen to have the UV cutoff $\Lambda_{\text{(Palatini)}}$. In the small
field limit one gets
\begin{eqnarray}
\mathcal{U}_g =  g^{2} A_{\mu} A^{\mu} \psi^2 + \dots
\end{eqnarray}
which is precisely the quadratic part of the vector boson Lagrangian. The
higher order terms are all suppressed by $\Lambda_{\text{(Palatini)}}$ as
can be guessed from (\ref{eq:a1-Pala-exact}).

Finally, the Yukawa contribution $\mathcal{U}_f$ in (\ref{eq:a2}) can be
expressed in terms of $\psi$ by using (\ref{rel-palatini}) so that
\begin{eqnarray}
\mathcal{U}_f = \Lambda_{\text{(Palatini)}} \tanh \left(\frac{\sqrt{\zeta}\psi}{M_{\text{Pl}}}\right) \overline{\Psi}\Psi
\end{eqnarray}
which equals
\begin{eqnarray}
\mathcal{U}_f =  \overline{\Psi}\Psi\psi + \dots
\end{eqnarray}
in the limit of small Higgs field $\psi$. The higher order terms buried in
$\dots$ are all suppressed by the inverse powers of $ \Lambda_{\text{(Palatini)}}$.

The detailed analysis of Higgs-Higgs, Higgs-gauge and Higgs-fermion interactions above makes it clear that the UV cutoff scale relevant for the UV embedding in the Palatini formalism is ${M_{\text{Pl}}}/{\sqrt{\zeta}}$.
This is larger than the corresponding scale in the metric formalism by a factor of $\sqrt{6\zeta}$. The UV scales
in the two formalism are shown comparatively in (\ref{UV-scales}).

Before we conclude let us briefly discuss the UV cutoff for large values of the Higgs field. For this purpose it is sufficient to expand the $\tanh(\sqrt{\zeta}\psi/M_\text{Pl})$ function with $\psi=\psi_0+h$ in powers of~$h$ in the limit of large~$\psi_0$,
\begin{eqnarray}
\tanh \left(\frac{\sqrt{\zeta}\psi}{M_{\text{Pl}}}\right)
= 1-\exp{\left(-2\frac{\sqrt{\zeta}\psi_0}{M_{\text{Pl}}}\right)}
\frac{2\exp{\left(-2\sqrt{\zeta}h/M_{\text{Pl}}\right)}}
{1+\exp{\left(-2\sqrt{\zeta}(\psi_0+h)/M_{\text{Pl}}\right)}}.
\end{eqnarray}
The last term of the right-hand side corresponds to a series of higher-order interactions in~$h$, which is suppressed during inflation by the tiny coefficient $\exp{\left(-2\sqrt{\zeta}\psi_0/M_{\text{Pl}}\right)\sim 10^{-12}}$ \cite{Palatini-HI-0803}. As a result, the effective UV cutoff is pushed from $\Lambda_\text{Palatini}$ to even higher energies.

Throughout the text,  the analyses have been exclusively restricted to the classical regime.  No mention of quantum corrections has been given. It is 
easy to see that, whichever formalism is used, a non-minimal Higgs-curvature coupling of the form $ g^{\mu \nu} {\mathbb{R}}_{\mu \nu}(\widehat{\Gamma}) {\mathcal{H}}^{\dagger} {\mathcal{H}}$ is radiatively induced. This happens just because of the coupling of the metric tensor to the Higgs sector, and generates a small non-minimal coupling~\cite{NonMinInfl-004-1002}. This correction thus gives rise to a mixture of the Metric and Palatini formalisms. Nevertheless, since successful inflation requires $\zeta$ to be very large, we safely neglect such renormalization effects\footnote{We thank the referee for pointing out this effect to us.}.

\section{Conclusions}
\label{sec:Conclusions}
We have explored unitarity violation in Higgs inflation by considering
the metric and Palatini formalisms of gravitation, in a comparative fashion.

In purely metrical gravity, where going from the Jordan frame action~(\ref{eq:Action-Jordan})
to the Einstein frame action~(\ref{eq:Action-Einstein}) proceeds with the
conformal transformation~(\ref{eq:ConfTrafo}), both Ricci tensor and Ricci scalar get
transformed and hence the dimension-6 operator in (\ref{eq:Action-Einstein}) appears. This
higher dimensional operator, for low values of the Higgs field $\phi$, violates unitarity
at a scale $\Lambda_{\text{(metric)}}$ which coincides with the Hubble rate. In other words, the
Higgs inflationary dynamics sits right atop the UV cutoff of the Higgs sector, and hence, it
is highly sensitive to `unknown physics' beyond $\Lambda_{\text{(metric)}}$. There have
been various proposals for restoring unitarity in Higgs inflation~\cite{unitarity-burgess,
UnitaryByHand-1005,Giudice2010-1010}. Notably, in~\cite{UnitaryByHand-1005}, a fine-tuned
counter term is added to the Jordan frame action for canceling the unitarity violating
term (proportional to $f$) in (\ref{eq:Action-Einstein}).

In Palatini gravity, thanks to the immunity of the Ricci tensor (not Ricci scalar) to
the conformal transformation~(\ref{eq:ConfTrafo}), the unitarity violating
dimension-6 operator is absent in  (\ref{eq:Action-Einstein}). Simply, $f=0$
therein. This property leads one at once to a natural (as proven in~\cite{Palatini-HI-0803})
Higgs inflation free from unitarity violation. Indeed, the Palatini setup does neither need 
a fine-tuned counter term nor a set of new scalars beyond the UV cutoff~\cite{Giudice2010-1010}.
The UV cutoff $\Lambda_{\text{(Palatini)}}$ is parametrically much larger than the UV cutoff
in the metric formalism. This feature is explicated in equation (\ref{UV-scales}). One notes that,
with $\zeta\sim 10^{10}$ in the Palatini formalism and $\zeta\sim 10^{4}$ in the metric variant,
actually the two UV cutoffs $\Lambda_{\text{(Palatini)}}$ and $\Lambda_{\text{(metric)}}$ turn
out to be numerically close to each other. However, while $\Lambda_{\text{(metric)}}$ coincides
with the Hubble parameter, $\Lambda_{\text{(Palatini)}}$ turns out to be some 5 orders of
magnitude larger than the Hubble rate. This implies that, new degrees of freedom
with masses around $m_{\text{UV}}\sim\Lambda_{\text{(Palatini)}}$ are too heavy to remain `integrated-in'
in the Higgs potential to influence the evolution of the inflationary epoch. To this end, Higgs inflation in the Palatini formalism turns out to be a UV-safe and minimal scenario for
describing inflation.

\subsection*{Acknowledgments}
The work of FB has been supported by DIUE/CUR Generalitat de Catalunya
under project 2009SGR502, by MEC and FEDER under project FPA2007-66665
and by the Consolider-Ingenio 2010 program CPAN CSD2007-00042. The
work of DAD has been partially supported by the T{\"U}B{\.I}TAK project
109T718. DAD is grateful to Cristiano Germani for fruitful discussions
on inflationary dynamics. He also thanks to Bar{\i}{\c s} Ate{\c s},
Oktay Do{\~g}ang{\"u}n, Canan D{\"u}zt{\"u}rk and Selin Soysal for discussions
on conformal transformations in non-Riemannian gravity.

\end{document}